\documentclass{article}
\usepackage[utf8]{inputenc}
\usepackage{amsmath,amssymb}
\usepackage{graphicx}
\usepackage[a4paper,top=3cm,bottom=2cm,left=3cm,right=3cm,marginparwidth=1.75cm]{geometry}

\title{Extragalactic cosmic rays diffusing \\from two populations of sources}
\author{Silvia Mollerach and Esteban Roulet\\
Centro At\'omico Bariloche, Comisi\'on Nacional de Energ\'\i a At\'omica\\ 
Consejo Nacional de Investigaciones Cient\'\i ficas y T\'ecnicas (CONICET)\\ 
Av. Bustillo 9500, R8402AGP, Bariloche, Argentina}
\date{}

\begin{document}

\maketitle
\begin{abstract}
We consider the possibility of explaining the observed spectrum and composition of the cosmic rays with energies above $10^{17}$~eV in terms of two different extragalactic populations of sources in the presence of a turbulent intergalactic magnetic field (including also a fading Galactic cosmic-ray component). The populations are considered to be the superposition of different nuclear species having rigidity dependent spectra. The first extragalactic population is dominant in the energy range $10^{17}$ to $10^{18}$~eV and consists of sources having a relatively large density ($> 10^{-3}$~Mpc$^{-3}$) and a steep spectrum. The second extragalactic population dominates the cosmic ray flux above a few EeV, it has a harder spectral slope and has a high-energy cutoff at a few $Z$~EeV (where $eZ$ is the associated cosmic ray charge). This population  has a lower density of sources ($<10^{-4}$~Mpc$^{-3}$), so that the typical intersource separation is larger than few tens of Mpc, being significantly affected by a magnetic horizon effect that strongly suppresses its flux for energies below $\sim Z$~EeV. We discuss how this scenario could be reconciled with the values of the cosmic-ray source spectral indices that are expected to result from the diffusive shock  acceleration mechanism.
\end{abstract}

\section{Introduction}

In spite of all the progress that has been achieved in the study of the high-energy cosmic rays, their sources still remain largely unknown. It is believed that for energies of at least that of the `knee' spectral steepening observed at $E_{\rm k}\simeq 4$~PeV \cite{ku59}, the cosmic rays (CRs) are predominantly of Galactic origin, possibly accelerated in supernova remnants or pulsars, while above the `ankle' spectral hardening observed at $E_{\rm a}\simeq 5$~EeV \cite{li63}, the CRs are expected to be predominantly of extragalactic origin, possibly accelerated in active galactic nuclei  or gamma ray bursts. The precise location of the transition between Galactic and extragalactic CRs is a matter of debate. Some scenarios associate it to the `second-knee' steepening of the spectrum observed at $E_{\rm sk}\simeq 100$~PeV \cite{berg07}, that would correspond to the break associated to the steepening of the heavy Fe component of the Galactic CRs in models where the knee would be the break associated to the lighter H/He Galactic component \cite{ca02,an05}. Other scenarios relate it to the ankle feature, associating it to the energy at which a harder extragalactic population would be overtaking the more steeply falling Galactic one.

Besides the spectral features, another important handle to understand the origin of the CRs is their composition, since changes in the average nuclear masses, as well as on the spread of their values, can provide clues about the source populations producing them. Indeed, the average composition is observed to become increasingly heavy from the knee up to the second-knee \cite{msu,yak,an05,ap13}, supporting scenarios in which the Galactic CRs get suppressed in a rigidity dependent way, so that the component of charge $eZ$ gets suppressed above an energy $ZE_{\rm k}$ \cite{pe61}. This suppression could either be due to an acceleration cutoff at the sources or, alternatively, be due to a more efficient diffusive escape from the Galaxy, since being both effects of magnetic nature they naturally depend on the particle's rigidities. The composition is observed to become lighter at EeV energies, suggesting the emergence of a new type of source population \cite{al14}, or eventually that a strong photodisintegration of heavy nuclei takes place at the sources,  producing large amounts of secondary protons at energies of a few EeV \cite{un15}. According to the Pierre Auger Observatory data, above the ankle energy the CRs appear to become increasingly heavy \cite{augerxm}, what possibly indicates that a rigidity dependent suppression is also present at the highest energies. Another relevant result is that the spread in the CR masses appears to become quite narrow  above the ankle, suggesting that the heavier species that dominate at the highest energies have to be strongly suppressed for decreasing energies, so as to avoid the simultaneous presence of light and heavy species at energies near $E_{\rm a}$. 

A final ingredient that should help to understand the CR origin is the anisotropy in the distribution of their arrival directions (for a recent review, see \cite{rpp}). In particular, near the knee energy a dipolar modulation in the equatorial component of the anisotropy has been observed by IceCube and IceTop \cite{ic12,ic16} that points close to the Galactic center direction, which is consistent with a predominant Galactic origin for the CRs at these energies. At higher energies, and up to $\sim 1$~EeV, the equatorial dipolar phases remain not far from the right ascension of the Galactic center, although the dipolar amplitudes are not significant \cite{lsra19}. The restrictive upper-bounds on the amplitudes, which are required to be below $\sim 1.5$\% in the range 1 to 4~EeV, combined with the observation that at these energies the composition is relatively light, disfavors a Galactic origin for this predominant light component, since if this were the case the anisotropy would be expected to be much larger \cite{lsl13}. At energies above 8~EeV, a significant dipolar anisotropy has been observed, pointing in the opposite hemisphere with respect to the Galactic center direction \cite{LSA17}, which is indicative of an extragalactic origin for the CRs at these energies. Moreover, some hints of more localized anisotropies, with hot spots on typical angular scales of 20$^\circ$ appearing at the highest energies, have been reported \cite{augersb,tahs} and, if confirmed, they may help to  identify the first sources of ultrahigh-energy cosmic rays (UHECRs).

An important observation is that the spectrum is dominated by heavy elements at the highest energies and that these elements are strongly suppressed for decreasing energies, so as to allow for the composition to become mostly light  near the ankle energy. This can be interpreted as resulting from the emission of different mass components having a rigidity dependent cutoff at energies of a few $Z$~EeV, that suppresses the light components above the ankle energy. Below this cutoff, the components need to have a very hard source spectrum so as to allow for the abrupt  emergence of the heavy components at the highest energies. In particular, assuming a power-law source differential spectrum  $\Phi(E)\propto E^{-\gamma}$ for each of the mass components, 
a fit to the Auger Observatory data on the spectrum and composition allows to determine $\gamma$ \cite{combfit}. The actual value of the common spectral index $\gamma$ turns out to depend on the hadronic model considered to describe the interactions in the atmosphere (as well as on other assumptions, such as the evolution of the sources or the extragalactic background light model). For instance,  for the EPOS-LHC model values of $\gamma<1.3$ are obtained, while for Sibyll 2.1 or QGSJET II-04 even harder spectra, with $\gamma<-1.5$, turn out to be preferred.
 These small required values are however in tension with the expectations from the CR diffusive shock acceleration scenarios, which typically predict that $\gamma\simeq 2$ to 2.4 (for a review see \cite{longair}). An alternative scenario was proposed in \cite{difu1}, where it was suggested  that the hard spectrum that has been inferred for the heavier mass components above a few EeV could be a consequence of the effects of the propagation of the CRs through the intervening extragalactic magnetic fields.  In particular, if the closest sources are at distances larger than few tens of Mpc, as the energy decreases below $Z$~EeV the propagation time of the diffusing CRs can become longer than the lifetime of the sources, and the CRs reaching the Earth would then be suppressed for decreasing energies due to the so-called magnetic horizon effect. For this suppression to be significant, the strength of the magnetic fields should be sizable ($B\gg {\rm nG}$)  and their coherence length should preferentially not be too large ($l _{\rm c}\ll{\rm Mpc}$). We note that the properties of the extragalactic magnetic fields are poorly known, being constrained indirectly from observed Faraday rotation measures of polarized sources, synchrotron  emission, etc. \cite{fe2012}, or being estimated alternatively from simulations of structure formation that include seed magnetic fields, from which a broad range of predictions are obtained \cite{dolag05,enzo17} (see \cite{widrow02,vallee04,vallee11} for reviews). Note that the presence of the Galactic magnetic field is not expected to affect significantly the spectrum and composition of the extragalactic flux component, and we will hence  ignore it.

In this work we consider a scenario that can account for the main features of the spectrum and composition measurements for all energies down to 100~PeV. It consists of two main extragalactic source populations contributing to the UHECRs, and a Galactic component which progressively fades away above 100~PeV and that contributes already less than $\sim 10$\% to the CR flux at 1~EeV. The extragalactic populations are considered  to arise from the superposition of five representative nuclear components at the sources: $i={\rm H}$, He, N, Si and Fe. They are assumed to originate from continuously emitting  sources with power-law CR spectra, $\Phi_i\propto f_i E^{-\gamma}$, with $f_i$ the fractional contribution to the spectrum at a given energy arising from the nuclei of type $i$. The spectrum of the CRs reaching the Earth is obtained taking into account propagation effects, due both to interactions with the radiation backgrounds and to magnetic deflections in the intervening extragalactic magnetic fields. For simplicity, we model the effects of a source acceleration cutoff directly by introducing a rigidity dependent exponential suppression in the fluxes reaching the Earth.

The first extragalactic population consists mostly of light nuclei (H, He and N) with a steeply falling source spectrum, with $\gamma\simeq 3.5$,  having a relatively large density of sources so as to lead to a typical intersource separation smaller than 10~Mpc (as is the case, for instance, for normal galaxies, starburst galaxies or Seyfert active galaxies). This population will dominate the CR flux in the range 0.1 to 2~EeV.
The second extragalactic population has instead a smaller source density (as could be the case,  for instance, for powerful radiogalaxies, blazars or galaxy clusters), so that the larger intersource separation leads, through a magnetic horizon effect caused by the CR deflections in the intergalactic magnetic fields, to a significant suppression of its flux for energies smaller than $\sim Z$~EeV, as was the case in the scenario suggested in \cite{difu1}. This population has significant amounts of heavier elements (He, N, Si and Fe), which also lead to large numbers of secondary protons through their photodisintegration,  and dominates the CR flux above a few EeV.

A somewhat similar two component scenario, but in which the high-energy CR flux originated from one (or few) nearby extragalactic powerful source emitting since relatively recent times, so that the magnetic horizon suppression could be sizable in spite of the relatively closer distance to the sources, was proposed in \cite{mr19}. In the discussion of the present scenario, that includes instead sources emitting since very early times,  we will consider different models for the cosmological evolution of the CR luminosities of the extragalactic populations. 

\section{Model for  the cosmic ray fluxes}

The total differential flux of cosmic rays with energies above 0.1~EeV will be  modelled with contributions coming from a Galactic population, $\Phi^{\rm G}$, and the two mentioned extragalactic populations: $\Phi^{{\rm XG}l}$, that is  dominant at low energies (between 0.1 and few EeV) and $\Phi^{{\rm XG}h}$, that is  dominant at high energies (above a few EeV), with
\begin{equation}
\Phi^{\rm tot} (E) = \Phi^{\rm G} (E) + \Phi^{{\rm  XG}l} (E) + \Phi^{{\rm XG}h} (E). 
\label{eq:phitot}
\end{equation}

The Galactic population is modelled, following \cite{mr19gal}, as a superposition of five nuclear components with relative fractions consistent with the direct measurements performed at $\sim 100$~TeV, and with rigidity dependent broken power laws with a high-energy exponential cutoff, with parameters determined from a fit to spectrum and composition data obtained between 1~PeV and 1~EeV. Since we are mostly interested in the extragalactic populations present at energies above 100~PeV, we keep the Galactic spectrum fixed in the analysis.

Each one of the extragalactic populations is modelled with five mass groups  plus the secondary nucleons that are produced during the propagation as a consequence of the interactions with the radiation backgrounds
\begin{equation}
\Phi^{{\rm XG}I} (E)= \sum_i \Phi^{ I}_i (E) + \Phi^{ I}_{\rm sp} (E),
\label{eq:phixg}
\end{equation}
where the sum runs over $i$ = H, He, N, Si and Fe,  for ${I} = l,h$. The source flux for each one of the mass group representative elements of the low or high extragalactic populations will be modelled as a power-law spectrum with spectral index $\gamma_{ I}$ up to a rigidity-dependent energy at which the acceleration at the sources is cut off, leading to an effective exponential suppression of the fluxes observed at the Earth above energies $Z_iE^{ I}_{\rm cut}$. 

The effects of the  interactions with the radiation backgrounds are taken into account by introducing a modification factor $\eta^i (E)$, defined as the ratio between the spectrum from a continuous distribution of  sources obtained including the attenuation effects and the spectrum that would have been expected from the same sources in the absence of interactions \cite{dip}.  The attenuation factors have been found to be quite insensitive to the source spectral index considered, although they depend on the cosmological evolution adopted for the luminosity of the sources. We will consider two representative cases of source evolution: a constant luminosity up to $z_{\rm max}=1$ (no evolution, NE) and a luminosity proportional to the star formation rate (SFR), for which we adopt the parametrization from \cite{ho06}, assuming that the source intensity evolves as $(1+z)^{3.44}$ up to redshift 0.97, evolving then as $(1+z)^{-0.26}$ for larger redshifts to then fall as $(1+z)^{-7.8}$ for redshifts above 4.48. These two illustrative cases bracket a wide range of plausible source evolution scenarios. 

We parametrize the attenuation factors for each of the mass groups considered following the approach of \cite{mr19}, and the parametrizations used are reported in the Appendix. One then has that, neglecting the possible effects associated to magnetic deflections and finite source distances,  
\begin{equation}
\Phi^{ I}_i (E) = \Phi^{ I}_0  f^{ I}_i\left(\frac{E}{\rm EeV}\right)^{-\gamma_{ I}} \eta^i(E)\frac{1}{\cosh(E/Z_iE^{ I}_{\rm cut})},
\label{phid}
\end{equation}
where the different fractions are defined at low enough energies such that the attenuation effects are negligible, and they satisfy $f^{ I}_{\rm H}+f^{ I}_{\rm He}+f^{ I}_{\rm N}+f^{ I}_{\rm Si}+f^{ I}_{\rm Fe}=1$ (equivalently, they can be considered as being the fractions in the source flux at an energy smaller than the H acceleration cutoff). Note that the cosh$^{-1}$  function allows to smoothly match the exponential suppression of the flux at energies higher than $ZE_{\rm cut}$ with the spectrum present at lower energies.

The secondary protons arise from the fragmentation of the different nuclei during propagation, and the resulting flux depends on the mass number, spectral index and source evolution of the component considered. They can be parametrized following the results of \cite{mr19}, and the parametrizations used are also reported in the Appendix.

The finite distance to the closest sources affects the attenuation of the high-energy population at the highest energies, and we include this effect by directly computing the expected attenuation for any adopted intersource separation, although in the scenarios considered  it is actually the source cutoff that provides the dominant attenuation effect at the highest energies. The combination of the finite source distance and the presence of intergalactic magnetic fields also determines the attenuation of the spectrum of the high-energy population for decreasing rigidities, as we now discuss. 

\section{The magnetic horizon effect}

One crucial ingredient  for the high-energy population of the present scenario is the spectral suppression appearing for decreasing  energies as a consequence of the magnetic horizon effect \cite{le04,be06,gl07}. This suppression  results from the combination of the relatively large intersource separation of this component and the diffusive propagation through the intergalactic magnetic fields, which implies that,  even for the closest sources, it may take longer than the age of the source for the low-energy CRs to reach the Earth. 
For the simple model of an isotropic turbulent magnetic field, characterized by an RMS strength $B$ and coherence length $l _{\rm c}$, the suppression can be accurately described through the analytic procedure developed in ref.~\cite{difu1}.\footnote{A fit to the Auger data above 5~EeV using non-uniform extragalactic magnetic field configurations was performed  in \cite{wi18}.} 

To obtain the suppression we compute, using the analytic solution developed by Berezinsky and Gazizov \cite{be06,be07} describing the difusion of CRs in an expanding Universe,  the spectrum of protons resulting from a distribution of sources with a given density, as well as that for a continuous distribution of sources, and obtain the ratio between them, that we call $G(E)$. Note that according to the  propagation theorem \cite{al04}, the total CR flux in the limit of a continuous distribution of sources should be the same as that obtained ignoring magnetic field effects. Then, the knowledge of the magnetic suppression factor $G(E)$ allows to account for the effects of the magnetic horizon just by multiplying the spectrum obtained in the absence of magnetic fields by $G(E)$. 
The suppression depends on  the average distance between sources, $d_{\rm s}$, and on the coherence length, $l_{\rm c}$, through the combination 
\begin{equation}
X _{\rm s}\equiv \frac{d _{\rm s}}{\sqrt{R _{\rm H} l _{\rm c}}}\simeq \frac{d _{\rm s}}{65\ \rm Mpc}\sqrt{\frac{\rm Mpc}{l _{\rm c}}},
\label{xs.eq}
\end{equation}
where the Hubble radius is $R_{\rm H}\equiv c/H_0\simeq 4.3$~Gpc.
The average separation between the UHECR sources, $d _{\rm s}$, is related to their density $n _{\rm s}$ through $d _{\rm s}\simeq n _{\rm s}^{-1/3}$. For example, $d _{\rm s}\simeq 10$~Mpc for $n _{\rm s}=10^{-3}$~Mpc$^{-3}$ while $d _{\rm s}\simeq 100$~Mpc for $n _{\rm s}=10^{-6}$~Mpc$^{-3}$, which spans most of the range of UHECR source densities usually considered. The magnetic suppression is computed considering a distribution of radial distances to the CR sources that follows the average distances to the closest sources in a homogeneous distribution  \cite{difu1} (in particular, the closest source lies in this case at a distance $r_1\simeq 0.55d _{\rm s}$). 
The magnetic suppression depends on the magnetic field amplitude through the critical energy $E _{\rm c}$, defined as the energy for which the effective Larmor radius, given by
$r _{\rm L}= E/ZeB \simeq 1.1 \,(E/{\rm EeV}) / (ZB_{{\rm nG}}) \ {\rm Mpc}$, is equal to the coherence length (with  $B_{{\rm nG}}\equiv B/{\rm nG}$). Then, requiring that   $r _{\rm L}(E _{\rm c})=l _{\rm c}$ one finds that 
 $E _{\rm c}\simeq 0.9 Z B_{{\rm nG}}(l _{\rm c}/{\rm Mpc})$~EeV. The analytic solution from \cite{be06} is a function of the diffusion length, which for a
turbulent magnetic field with a Kolmogorov spectrum  can be accurately parametrized as
\cite{difu2} 
\begin{equation}
l_{\rm D}(E) = l _{\rm c}\left[ 4\left(\frac{E}{E _{\rm c}}\right)^2 + 0.9\left(\frac{E}{E _{\rm c}}\right) + 0.23\left(\frac{E}{E _{\rm c}}\right)^{1/3}\right].
\label{geldnew.eq}
\end{equation}
Note that the diffusion length is the typical distance after which a charged particle would be deflected by about 1~rad. 

\begin{figure}[h]
\centering
\includegraphics[scale=0.48,angle=270]{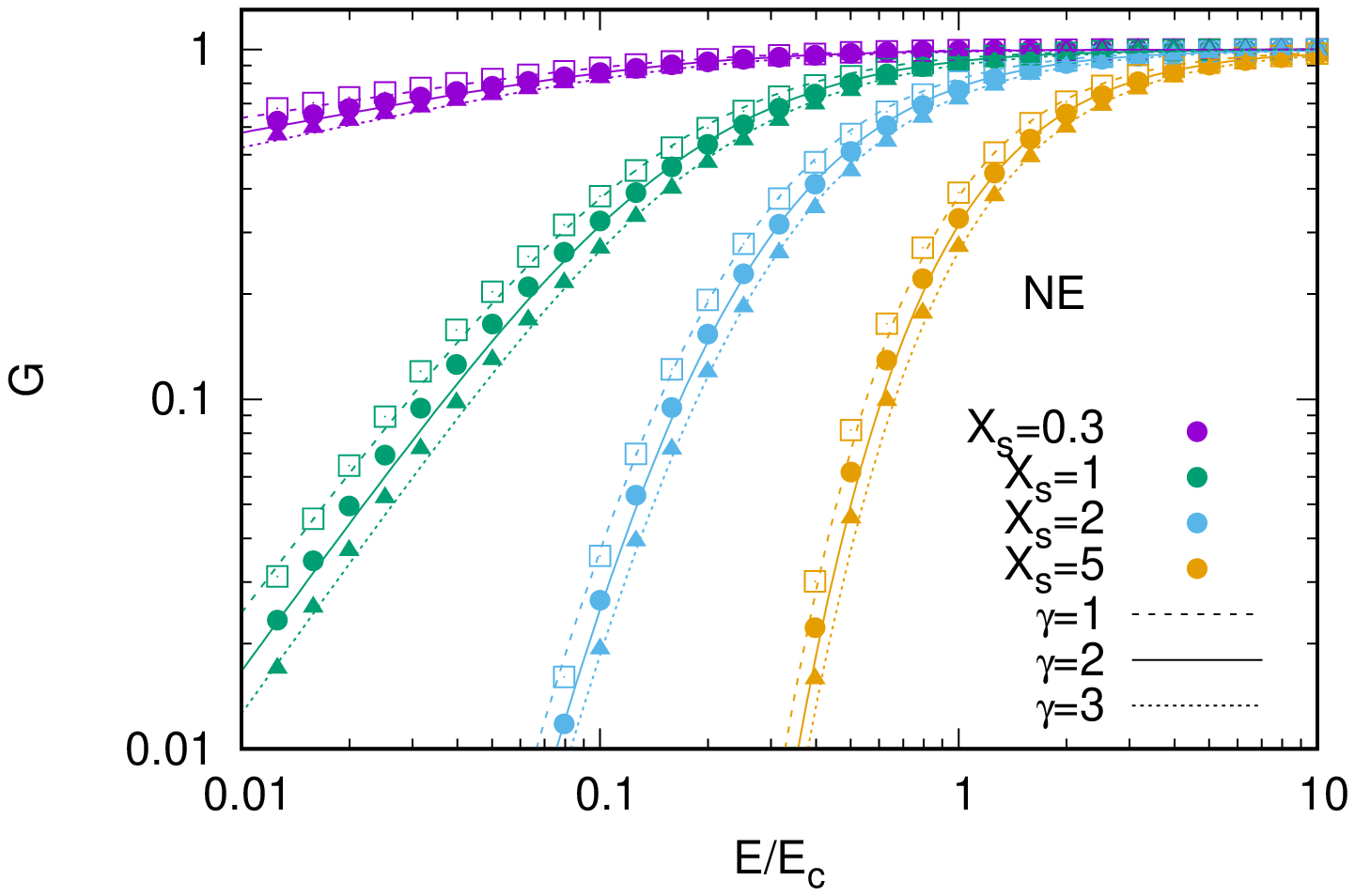}
\includegraphics[scale=0.48,angle=270]{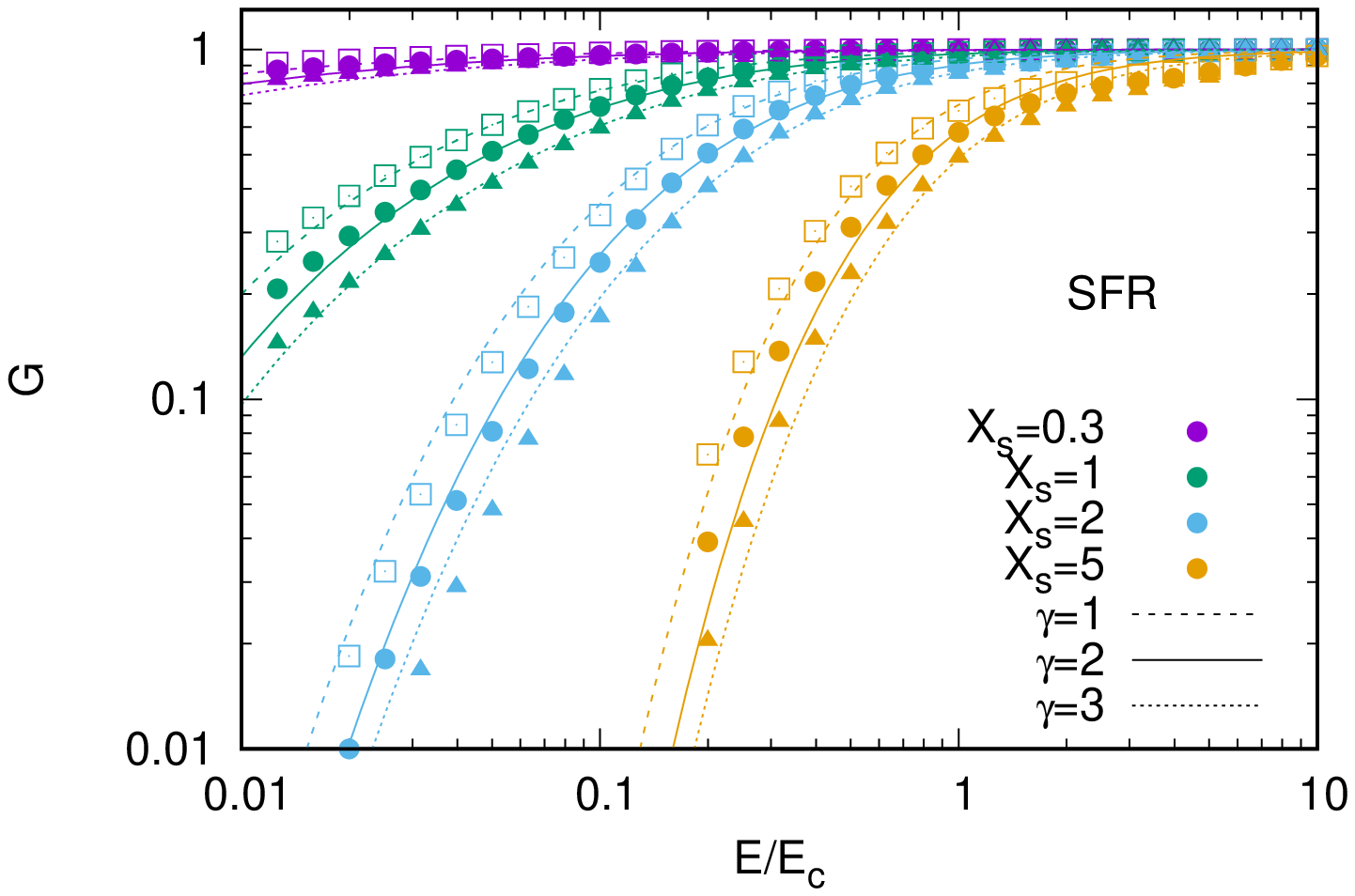}
\caption{Suppression factor $G(E/E _{\rm c})$ for different source evolution models, spectral index $\gamma$ and $X _{\rm s}$ parameter. The points are the results of the numerical computation while the lines correspond to the fits obtained using eq.~(\ref{gfit.eq}).}
\label{fig:zmev}
\end{figure}

The magnetic suppression turns out to also depend on the evolution of the luminosity of the sources with redshift. 
In Figure~\ref{fig:zmev} we show with points the suppression  factor $G$ obtained as a function of $E/E _{\rm c}$, for two models for the source evolution (NE and SFR) and for four different values of the mean source separation, corresponding to $X _{\rm s} = 0.3, 1, 2$ and 5.  The results in the plots for the SFR actually include sources just up to a maximum redshift of four, since the contribution from sources farther away is negligible. The magnetic suppression is stronger for larger intersource distance $d _{\rm s}$ (larger $X _{\rm s}$, lower density), as expected.  The suppression is weaker in the SFR evolution case, since the particles travelling for longer times, and thus reaching us from farther away, get more weight in this case. The suppression has also a slight dependence on the spectral index $\gamma$, and we display the results for $\gamma =1,~2$ and 3.
    
A good fit to the suppression factor can be obtained through the expression 
\begin{equation}
G(x)=\exp\left[-\left(\frac{a\,X _{\rm s}}{x+b (x/a)^\beta}\right)^\alpha\right],
\label{gfit.eq}
\end{equation}
with $x=E/E _{\rm c}$. This expression slightly improves the one adopted in ref.~\cite{difu1}, where a less accurate expression for $l_{\rm D}$ was used. 
The results of the fits, obtained using the parameters reported in Table~\ref{tab:gpar}, are shown as lines in Figure \ref{fig:zmev}. These fits are quite accurate for the different cases of source evolution and densities studied, and hence we will use them in the combined fit of the spectrum and composition data since they allow to consider different magnetic field parameters and source models without the need of performing new computations for each case.    

 \begin{table}[ht]
\centering
\begin{tabular}{c c c c c}
\hline\hline
    Evolution & $a(\gamma$) & $b(\gamma$) & $\alpha$($\gamma$)& $\beta$($\gamma$)\\
\hline
 NE & 0.206+0.026\ $\gamma$ &  0.146+0.004\ $\gamma$  & 1.83 - 0.08\ $\gamma$  & 0.13 \\
 SFR &0.135+0.040$\ \gamma$ & 0.254+0.040\ $\gamma$  & 2.03 - 0.11\ $\gamma$  & 0.29 \\
 
\hline
\end{tabular}
\caption{Parameters of the fit to the suppression factor $G(E/E _{\rm c})$ for the two models  of source evolution, as a function of the source spectral index $\gamma$.}
\label{tab:gpar}
\end{table}

We note that the magnetic suppression factor $G$ was obtained ignoring interactions during propagation, just keeping redshift effects, since this suppression is relevant only at  energies smaller than about $Z$~EeV, while the interactions are relevant mostly at higher energies. If one were to consider values of the parameters for which the magnetic suppression would appear at higher energies, the interactions could in principle also affect the magnetic suppression, as discussed in \cite{difu1}.

\section{The two extragalactic source population scenarios}

In this section we obtain the main features of the two extragalactic populations, as well as of the intergalactic magnetic fields, which are required in order that they lead to predictions in reasonable agreement with the observed CR spectrum and composition. We consider the measurements performed by the Pierre Auger Observatory for energies above 0.1~EeV. Data from other experiments exist in this energy range, but we do not include them since they rely on significantly smaller number of events and hence they should not significantly affect the results obtained. Moreover, different datasets are affected by different systematic uncertainties, such as those related to the different energy calibrations of each experiment, and this would further complicate a combined analysis. 
For the Galactic CRs, we will adopt the fluxes already derived in \cite{mr19gal} in a fit including lower energy data  (we just rescale the energy parameters of the fit in \cite{mr19gal}, which relied on the energy scale of the Telescope Array experiment, to the energy scale of the Auger experiment). 

We will fit the parameters describing the two extragalactic populations to the Auger spectrum data above 0.1~EeV from \cite{ve19,co19} as well as to the composition data obtained for $E\geq 0.16$~EeV in \cite{augerxm}. This last includes the derived values of the average logarithm of the mass number of the CRs, $\langle {\rm ln} A\rangle$, and its variance, $ \sigma^2({\rm ln} A)$. These are obtained from the measurements of the depth of maximum development of the air showers, $X_{\rm max}$, performed with the fluorescence detectors.
The relation between $\langle X_{\rm max}\rangle$ and $\langle {\rm ln}A\rangle$ depends on the hadronic model considered to simulate the CR interactions in the atmosphere and, for definiteness, we adopt in our analysis the results based on the Sibyll~2.3c model \cite{sibyll}, which leads to an inferred composition slightly heavier than that based on the EPOS-LHC model \cite{epos}. For consistency, the Galactic population that we adopt is also that obtained using the Sibyll~2.3c hadronic model in \cite{mr19}, and we consider the scenario including a high-energy cutoff for this population. 

 \begin{figure}[h]
\centering
\includegraphics[scale=.72,angle=0]{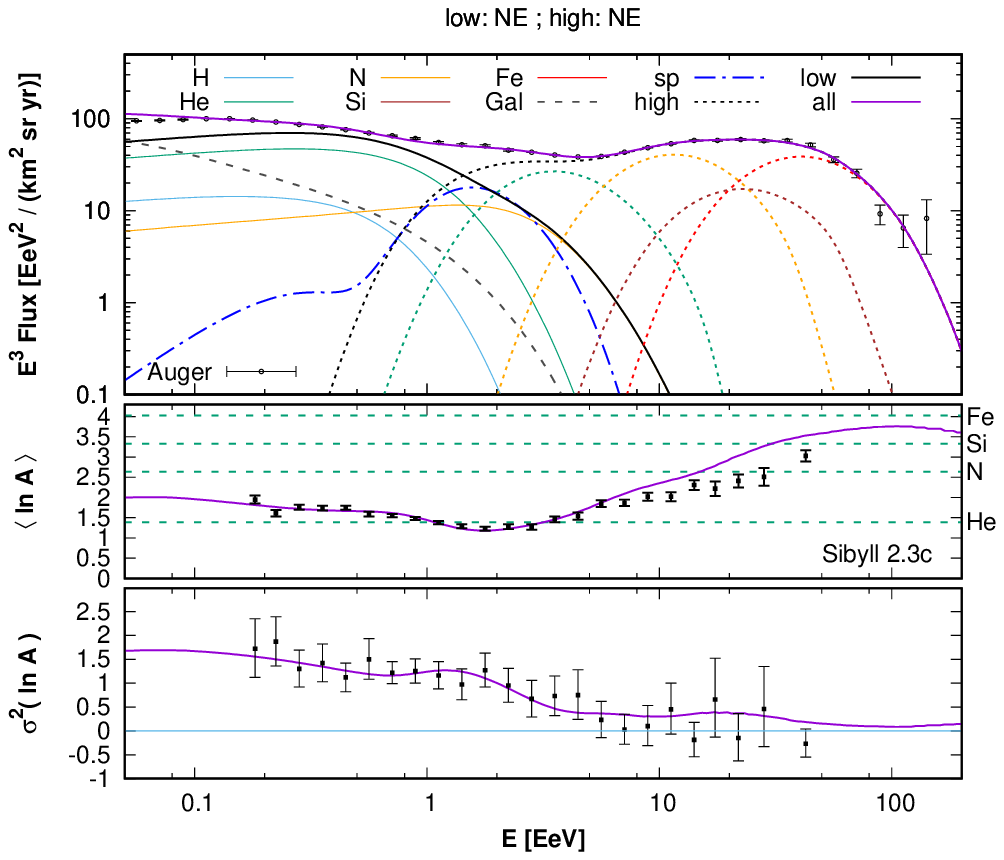}
\includegraphics[scale=.72,angle=0]{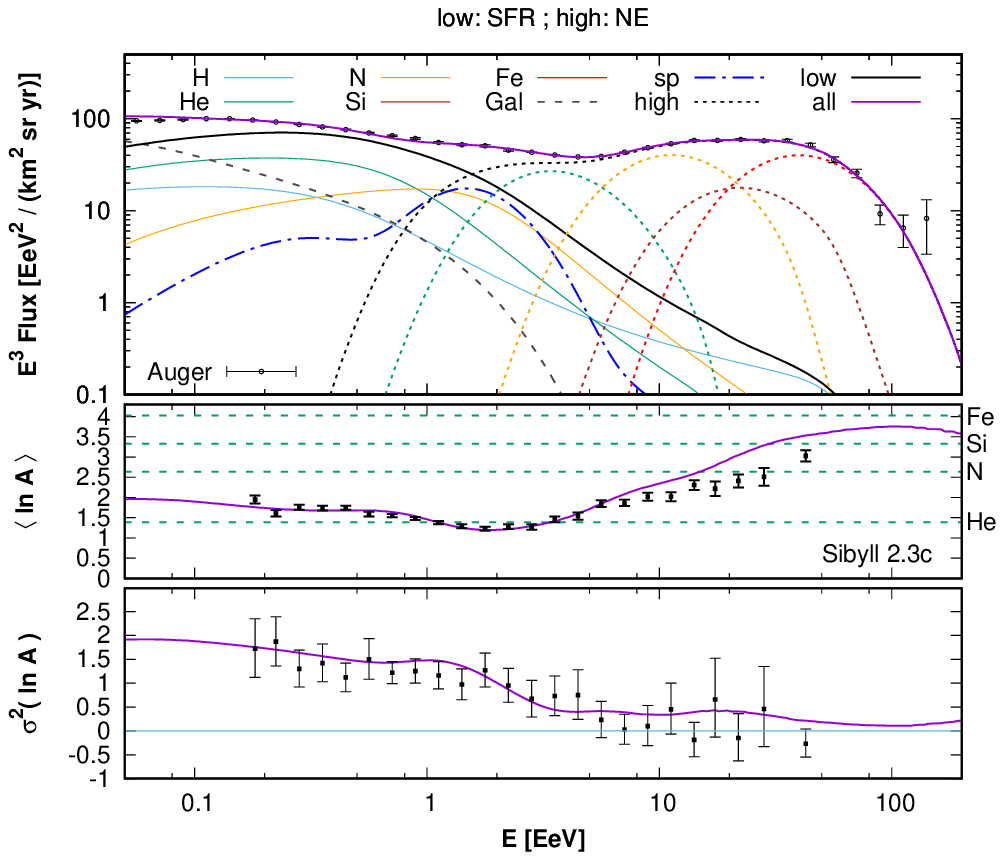}
\includegraphics[scale=.72,angle=0]{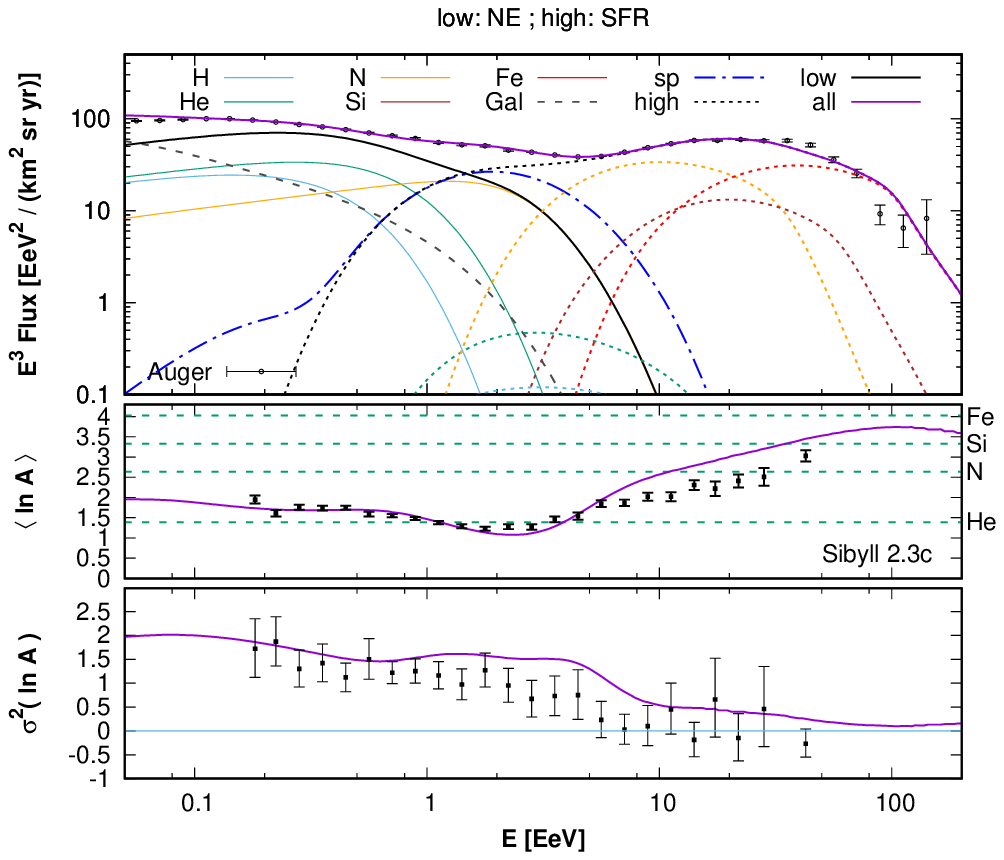}
\includegraphics[scale=.72,angle=0]{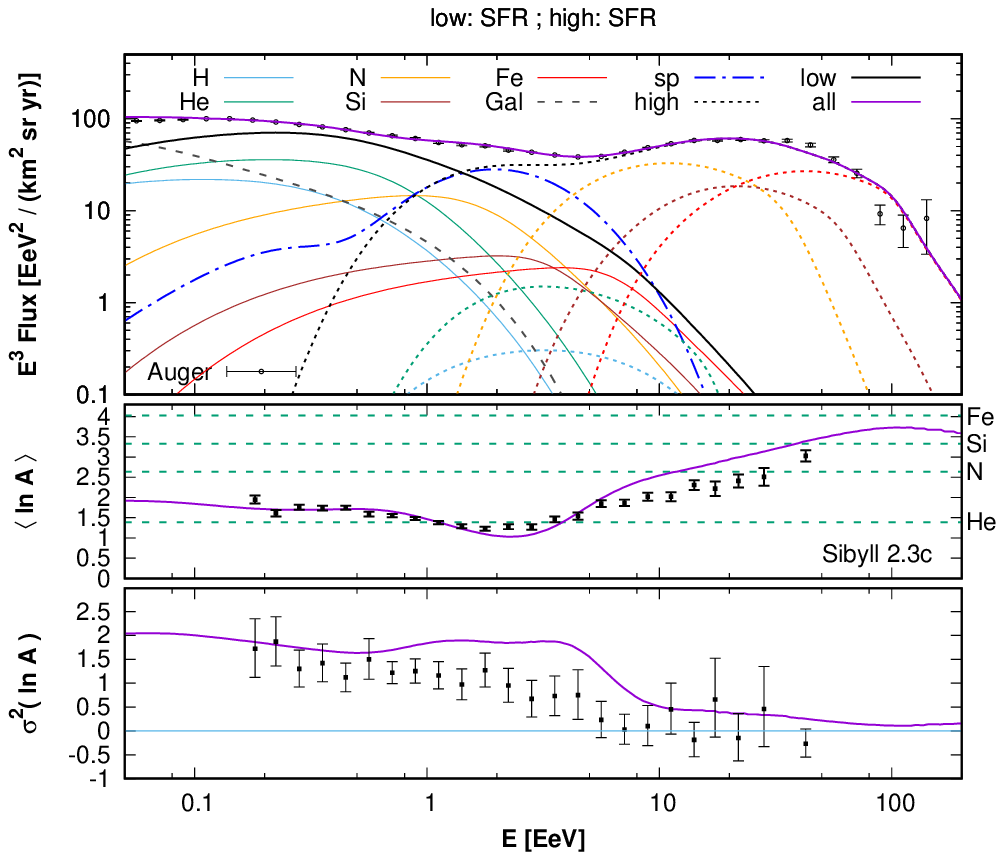}
\caption{Spectrum and composition for different assumptions on the cosmological evolution of the luminosity of the two extragalactic populations, adopting $E _{\rm c}=2$~EeV and $d _{\rm s}^h=75$~Mpc. We show separately the contributions to the spectrum from the different mass groups of the low (continuous lines) and high (short dashes) extragalactic populations, as well as their total contributions in black. The total contribution of the secondary protons from both populations is indicated with blue dot-dashed lines, the Galactic contribution with black long-dashed lines and the total spectrum is displayed as the violet continuous line.}
\label{fig:fits}
\end{figure}

In Fig.~\ref{fig:fits} we display the results obtained for the spectrum, $\langle {\rm ln}A\rangle$ and $\sigma^2({\rm ln} A)$, making different assumptions for the cosmological evolution of the luminosity of the two extragalactic source  populations (either with no evolution, NE,  or assuming an evolution that follows the star formation rate, SFR). We adopted in the plots a critical energy $E _{\rm c}=2$~EeV to characterize the effects of the extragalactic magnetic field, and an intersource separation for the high-energy population  $d _{\rm s}^h=75$~Mpc to evaluate the attenuation at the highest energies. In Table~\ref{tab:fitpars} we report the values of the different parameters that are obtained  in each case through the minimization of the $\chi^2$ function constructed considering the statistical uncertainties of the different measurements. One can see from the figure that the overall agreement  of the models with the data points is quite good for all the energy range considered.  In the spectrum plot we show separately the contribution of the different mass components for each extragalactic population. One should keep in mind that,  for instance, the component labelled as Si includes all the leading nuclear fragments arriving to the Earth that were produced in the photodisintegration of the  nuclei emitted as Si at the source, and the secondary protons resulting from the interactions of all nuclear species are displayed separately. The lowest energy bump in the flux of secondary protons is mostly due to the low-energy extragalactic population, while the larger bump appearing at higher energies is mostly due to the high-energy extragalactic population.

\begin{table}[t]
\centering
\caption{Parameters  obtained in the fit adopting $E _{\rm c}=2$~EeV and $d _{\rm s}^h=75$~Mpc. The first column indicates the evolutions assumed for the low and high-energy extragalactic populations respectively. }
\bigskip
\begin{tabular}{c c c c c c c c c c}
\hline\hline
  Evolution &  $\gamma _{\rm l}$ &$E_{\rm cut}^l$ [EeV] & $X _{\rm s}^l$  & $f _{\rm H}^l$ & $f_{He}^l$ & $f_N^l$ & $f_{Si}^l$& $f_{Fe}^l$ & $\phi_0^l$ [1/km$^2$\,yr\,sr\,EeV]\\
 \hline
NE-NE & 3.5 & 0.44& 0.63 & 0.13 & 0.63 & 0.24 & 0 & 0 & 101 \\
SFR-NE & 3.4 & 100 & 0.79 & 0.19 & 0.51 & 0.30 & 0 & 0 & 77 \\
NE-SFR & 3.5 & 0.30 & 0.70 & 0.20 & 0.40 & 0.40 & 0 & 0 & 140 \\
SFR-SFR & 3.5 & 1.2 & 0.95 & 0.17 & 0.41 & 0.26 & 0.08 & 0.08 & 93 \\
\hline\hline
  Evolution &  $\gamma _{\rm h}$ &$E_{\rm cut}^h$ [EeV] & $X _{\rm s}^h$ & $f _{\rm H}^h$ & $f_{He}^h$ & $f_N^h$ & $f_{Si}^h$& $f_{Fe}^h$ & $\phi_0^h$ 
  [1/km$^2$\,yr\,sr\,EeV]  \\
 \hline
NE-NE & 2.0 & 1.6& 3.6 & 0 & 0.52 & 0.31 & 0.07 & 0.10 & 196 \\
SFR-NE & 2.0 & 1.4 & 3.7 & 0 & 0.52 & 0.30 & 0.07 & 0.11 & 221 \\
NE-SFR & 2.4 & 5.3 & 5.2 & 0 & 0.01 & 0.55 & 0.15 & 0.29 & 873 \\
SFR-SFR & 2..4 & 5.0 & 5.5 & 0 & 0.03 & 0.52 & 0.20 & 0.25 & 1000 \\
\end{tabular}
\label{tab:fitpars}
\end{table}

There are several salient features which are common to all the different scenarios. In particular, between 0.1 and $\sim 2$~EeV the spectrum is dominated by the light component (H, He and N) of the low-energy population and this population has negligible contributions from heavier elements.\footnote{Given that the Si and Fe components of the low-energy population cannot be reliably constrained separately, we just considered in the fits equal fractions for both of them.} The lack of heavy elements in this component helps to reduce the spread in mass values, leading to a good agreement with the variance of ln$A$ that is observed. In the energy range between 1 and 5~EeV, the main contributions are from the N of the low-energy population  as well as a significant amount of secondary protons from the high-energy population. Above the ankle energy, the main contributions are those from the N, Si and Fe components of the high-energy population, with the larger masses progressively dominating for increasing energies.\footnote{Note that the average CR masses that are predicted by the models above $\sim 10$~EeV are slightly heavier than the values inferred from the data. This conclusion depends however on the hadronic model being considered, and given that at these energies one needs to rely on extrapolations of the hadronic models beyond the energies at which they are constrained by colliders, significant systematic uncertainties could still  affect the values of $\langle {\rm ln} A\rangle$ that are inferred from observations in this energy range. Moreover, we did not consider the impact  of the experimental systematic uncertainties that affect  the determination of the depth of shower maximum $X_{\rm max}$ as well as the energy scale,  which could also affect the average mass that is inferred from the data.}
  The  low-energy population ended up having a very steep spectrum, with $\gamma_l\simeq 3.5$. Since this spectral index is mostly determined by the shape of the spectrum in the decade below 1~EeV, it has almost no sensitivity to the source evolution adopted for the low-energy population. 
   When the low-energy population has no evolution, one generally finds that its cutoff has a small value, $E_{\rm cut}^l<1$~EeV. When the evolution follows the SFR, which already leads to a steeper final spectrum due to the effects of the interactions which get enhanced at high redshifts, the resulting cutoff can be much larger,  even reaching the maximum value that we allowed of 100~EeV. However, the $\chi^2$ function has very little sensitivity to this parameter since above 20~EeV the low-energy population contributes already less than 1\% to the total flux.  Note that, in this kind of scenarios, the presence of a subdominant population of light CRs possibly extending up to the highest energies could prove  helpful in the attempts to identify some of the nearby sources through anisotropy studies. 

Regarding the  spectrum of the high-energy population, we are particularly interested in an explanation in which a source spectral index compatible with the expectations from diffuse shock acceleration gets effectively hardened by the magnetic horizon effects after the propagation is taken into account. We will hence just consider values for $\gamma_h$ in the range 2 to 2.4.
For the NE case, the spectral index obtained tended to the lower boundary of the range considered, $\gamma_h\simeq 2$, with the cutoff energy having  typical values of about 1.5~EeV. In this case an even harder spectrum ($\gamma\simeq 1.2$) would have been preferred by the fit, but with only slight improvements in the $\chi^2$ value, with a correlated reduction of $X _{\rm s}^h$ and a decrease in $E_{\rm cut}^h$. Since the modelling of the extragalactic populations that we consider is very simplistic, with just five different components of uniformly spaced equal intensity sources with similar spectra, and there are also possible unaccounted systematic effects related to the assumptions about the hadronic models, the energy calibration, etc., we favor in our analysis the possibility of getting a source spectral index closer to the expectations from diffusive shock acceleration ($\gamma\geq 2$) rather than to strictly minimizing the $\chi^2$ function by allowing less physically motivated regions of the parameter space. In the case of the SFR evolution, the spectral slope of the  high-energy population turns out to be  $\gamma_h\simeq 2.4$, and the cutoff energies have  typical values of about $\sim 5$~EeV. The values obtained for the cutoff energy of the high-energy population are essential in order to ensure that the light component of this population does not extend much beyond the ankle energy. Let us note that the global $\chi^2$ value per degree of freedom obtained in the fits turns out to be smaller for the cases in which the high-energy population has no evolution  ($\chi^2/{\rm dof}\simeq 4$) than for the cases with an evolution following the SFR ($\chi^2/{\rm dof}\simeq 6$).

One can see from the plots in Fig.~\ref{fig:fits} that the models that consider a high-energy population with a SFR evolution lead to a larger amount of secondary protons at a few EeV energies, having also a broader distribution.  On the other hand, when the high-energy population has no cosmological evolution, the amount of secondary protons gets reduced and an increased He contribution from the high-energy population is then required.

The parameter $X _{\rm s}$ determining, together with $E _{\rm c}$,  the magnetic horizon effect, needs to be much larger for the high-energy population than for the low-energy one, since this suppression is crucial to lead to an effectively  very hard spectrum for each of the mass components of the high-energy population. This can naturally result if the high-energy population has a much lower source density than the low-energy one. One typically obtains, for the initially adopted value of $E _{\rm c}=2$~EeV, that $X _{\rm s}^h\simeq 3.6$ in the no evolution case and $X _{\rm s}^h\simeq 5$ for the SFR case, while in all cases $X _{\rm s}^l<1$. Given that the required intersource separation would be $d _{\rm s}\simeq 65\,{\rm Mpc}X _{\rm s}\sqrt{l _{\rm c}/{\rm Mpc}}$, if we also require that $d _{\rm s}<100$~Mpc in order  that the high-energy sources are not too rare and not too suppressed at the highest energies by interactions during propagation, one concludes that the coherence length of the magnetic field should be of the order of galactic scales ($<100$~kpc) rather than of the order of the typical distance between galaxies ($\sim {\rm Mpc}$). 
On the other hand,  requiring that $l _{\rm c}>10$~kpc one would conclude that $d _{\rm s}^h>20$~Mpc for the NE case (while  $d _{\rm s}^h>40$~Mpc for the SFR case). This would imply a source density smaller than $10^{-4}$~Mpc$^{-3}$ ($10^{-5}$~Mpc$^{-3}$ respectively) for the high-energy population.

If we were to consider a different value of the critical energy, the main impact on the results would be that the preferred value of $X _{\rm s}$ would become smaller  for increasing values of $E _{\rm c}$. For instance, for the SFR-NE scenario one gets $X _{\rm s}^h\simeq 7.6$, 5.9, 3.7 and 1.7 for $E _{\rm c}=0.5$, 1, 2 and 10~EeV respectively. Given that $E _{\rm c}\simeq 0.9B_{\rm nG}(l _{\rm c}/{\rm Mpc})$, one finds that the required value of the extragalactic magnetic field needs to be sizable, of order $B\simeq 20\,{\rm nG}(E _{\rm c}/{\rm EeV})(50\,{\rm kpc}/l _{\rm c})$. Such large values of the extragalactic magnetic fields could result, for instance, from the amplification of primordial seeds \cite{enzo17}.

For the low-energy population, the values obtained of $X _{\rm s}^l\simeq 0.6$ to 1 suggest that the associated source density should be much larger, with $n _{\rm s}^l>10^{-3}$~Mpc$^{-3}$. The magnetic horizon suppression of the flux from this population should be important in shaping its spectrum at energies below 0.1~EeV. In this respect, the study of the low-energy ankle feature present at $\sim 20$~PeV could be helpful to further constrain $X _{\rm s}^l$ \cite{mr19gal}.

\section{On the steepness of the low-energy population spectra}

One property that was derived in the previous analysis is that the low-energy population needs to have, below its cutoff value,  a very steep spectrum with $\gamma\simeq 3.5$. This is significantly larger than the values 2 to 2.4 which are typically obtained in scenarios of diffusive shock acceleration. A possible way to obtain an effectively steeper spectrum from sources having a hard spectrum, but having a power-law distribution of values of the source  cutoff energies, was suggested in \cite{ka06}, and we here comment on this alternative. 

Let us consider a population of  continuously distributed  sources having similar luminosities below their cutoff energies, with a common spectral index $\gamma _{\rm s}$ but having a distribution of cutoff energies. For simplicity we here assume the cutoff to be sharp, so that for any given source with cutoff energy $E_{\rm cut}$ the number of CRs emitted per unit time is  $q(E,E_{\rm cut})\propto  E^{-\gamma _{\rm s}}\Theta(E_{\rm cut}-E)$, with $\Theta$ the Heaviside function. Considering the cutoff values of different sources to have a power-law distribution  such that the source density satisfies d$n _{\rm s}(E_{\rm cut})/{\rm d}E_{\rm cut}\propto E_{\rm cut}^{-\beta}$, one would get, ignoring evolution and propagation effects, that the total flux at the Earth will be
\begin{equation}
    \Phi(E)\propto \int_E^\infty {\rm d}E_{\rm cut}\frac{{\rm d}n _{\rm s}(E_{\rm cut})}{{\rm d}E_{\rm cut}}q(E,E_{\rm cut})\propto E^{-\gamma _{\rm s}-\beta+1}.
\end{equation}
In this case, the spectrum resulting from the superposition of all the sources will have an effective spectral index $\gamma=\gamma _{\rm s}+\beta-1$. Hence, a steep spectrum with $\gamma\simeq 3.5$ could result, for instance, from $\gamma _{\rm s}=2$ if one considers $\beta\simeq 2.5$. 
If the sources have an evolution with redshift, 
the same reasoning can be applied to the emissivity from any redshift interval to conclude that
  it is equivalent to have  a population of sources with a steep spectrum $\gamma$  having all a large cutoff energy  or to have instead sources with a harder spectral index $\gamma _{\rm s}$ but  having a power-law distribution of cutoff energies, with $\beta=\gamma-\gamma _{\rm s}+1$. Note that  if  $E_{\rm cut}$ were to depend on redshift, this would   ultimately also modify  the effective source evolution of the model.

\section{Two populations with a common composition?}
In this section we consider whether the two extragalactic populations could be associated with a similar underlying composition, in such a way that the fraction of the different elements that are present in the medium in which the CRs get accelerated is similar for both populations. Even if this were the case, their spectral indices and cutoff energies could end up being different due to the different properties of the acceleration process involved in each case. 

If we denote as $f_i^0$ the fraction of the element $i$ that is present in the medium in which the acceleration takes place, and consider that all elements get fully ionized and are accelerated in a rigidity dependent way, one should expect then that the final cumulative source fluxes  above a certain threshold rigidity value should also have the same relative abundances, i.e.
\begin{equation}
 \frac{\int_{Z_iE_{\rm th}}^\infty {\rm d}E\,\Phi_i^s(E)}{\int_{E_{\rm th}}^\infty {\rm d}E\,\Phi _{\rm H}^s(E)}= \frac{f_i^0}{f _{\rm H}^0}.
\end{equation}
In particular, for a power-law source spectrum such that $\Phi_i^s(E)\propto f_iE^{-\gamma}$ (note that the fractions can be defined at the energy $E_{\rm th}\ll E_{\rm cut}$, and hence the effects of the source cutoff can be neglected here), this would lead to
\begin{equation}
  f_i\simeq f _{\rm H} \,Z_i^{\gamma-1}\,f_i^0/f _{\rm H}^0.
\end{equation}
If the low-energy and high-energy extragalactic populations were to originate from environments with similar composition fractions $f_i^0$, and the CRs were accelerated such that they end up having power-law spectra characterised by indices $\gamma_l$ and $\gamma_h$, one should then expect that
\begin{equation}
  f_i^l\simeq f_i^hZ_i^{\gamma_l-\gamma_h}.
\end{equation}
This implies that the composition of the accelerated CRs of the population with steeper spectrum should be enhanced  in heavier elements with respect to the population having a harder spectrum. This is however at odds with the results we obtained previously for the two extragalactic population scenarios considered, which indicated that the steeper low-energy population  had however a smaller fraction of heavier elements than the high-energy population. This then suggests that the CRs from the two populations get accelerated in environments having quite different distributions of elements (or, alternatively, that the heavy nuclei in the low-energy population get largely disintegrated during their acceleration). We also note that the compositions inferred for these two populations differ from the composition of the Galactic cosmic rays measured at lower energies. For instance, at $10^{14}$~eV, where $\gamma\simeq 2.7$, the composition of the different mass groups is $f_{\rm H}\simeq f_{\rm He}\simeq 0.35$ and $f_{\rm N}\simeq f_{\rm Si}\simeq f_{\rm Fe}\simeq 0.1$, which suggests that the nature of the sources responsible for these populations is different.

\section{Discussion}

We have considered a scenario in which the UHECRs are mostly extragalactic and arise from two main populations having different source densities, compositions, spectral indices and cutoff values. In these scenarios, the Galactic-extragalactic transition would take place slightly below the second-knee energy, with the low-energy extragalactic population dominating the CR spectrum in the range from $\sim 0.07$~EeV up to about 2~EeV while the high-energy population would dominate the spectrum at higher energies. One of the main features that was derived \cite{combfit} from the  spectrum and composition  inferred from the Auger Observatory measurements, is the requirement that the different components observed above the ankle energy need to have a very hard spectrum and that they also need to have a rigidity dependent source cutoff at energies of about a few $Z$~EeV. Instead of getting the hard spectrum as a result of a very hard injection spectrum at the source, in tension with the expectations from diffuse shock acceleration, we here considered the possibility that this be the result of the hardening produced during the propagation as a consequence of a magnetic horizon effect, as originally suggested in \cite{difu1}.\footnote{Yet another possibility to implement the magnetic horizon effect that suppresses the observed flux at low rigidities would be in scenarios in which  the high-energy sources are located in the cores of galaxy clusters \cite{ha16} since, given the  magnetic fields with typical $\mu$G strengths present in the cluster environments,  the confinement times of the charged CRs inside the clusters could be longer than  the times required for their subsequent propagation up to the Earth.} We here also combined this high-energy population with another extragalactic population dominating the flux below a few EeV, as had been considered in \cite{mr19} in a scenario in which the high-energy flux originated from nearby extragalactic sources within the Local Supercluster that were active since relatively recent times. In the scenarios considered in the present work, with continuous emission since the earliest times, the source density of the high-energy population needs to be small, typically $n _{\rm s}^h< 10^{-4} {\rm Mpc}^{-3}$, in order that the magnetic suppression be significant at energies $\sim Z$~EeV for acceptable values of the extragalactic magnetic field strength and coherence length. We generally obtain that the low-energy population has a small contribution from the elements heavier than N, while the high-energy population has a small contribution from H at the sources, although an important contribution of secondary protons at energies of a few EeV results from the photodisintegration of the heavy elements during their propagation. Since these protons are expected to be produced mostly at high redshifts, their flux would be  quite isotropic,  and hence one would expect that they tend to suppress the CR anisotropies at energies of a few EeV, in line with the present restrictive upper limits on the equatorial dipole amplitude, that should be below 1.5\% in the energy range 1 to 4~EeV \cite{lsra19}. We note that a difference with respect to the scenario in which the high-energy population  is due to a nearby source emitting since recent times would be the lack of significant amounts of secondary protons  in this last case \cite{mr19}. This kind of scenario then needs to include instead a larger fraction of light elements produced directly at the nearby source, which tends to enhance the predicted anisotropies, and this could help to distinguish between the different possibilities. A detailed study of these predictions would need to consider also the effects of the Galactic magnetic fields on the anisotropies.

The inferred source properties for the two extragalactic populations considered in this work depend significantly on the assumed source evolution, and hence a detailed determination of the CR composition could also help  to obtain information about the evolution of the sources. We note that the inferred source spectrum of the low-energy population turns out to be quite steep  and, as we mentioned, this could be an effective slope resulting from the combination of many harder sources  having a distribution of cutoff energies. This is clearly a very natural possibility, since the  cutoff energies will ultimately depend on the power of the sources and on the magnetic fields present in them, and  there is no reason for these quantities to be the same for all UHECR sources.

\section*{Appendix: Attenuation factors}

 We report here the attenuation factors $\eta$, both for protons and for the four representative heavier nuclear species considered in this work. They are given by the ratio between the spectrum of the particles reaching the Earth from a continuous (i.e. high density) distribution of sources including the attenuation effects with respect to the spectrum that would have been expected from the same sources in the absence of interactions. 
 
 Protons lose energy mainly through pair production and photo-pion production when interacting with the cosmic microwave background (CMB) radiation. The nuclei are affected by their photodisintegration off the photon backgrounds (which reduces the mass of the leading fragment and leads to the emission of secondary nucleons), as well as by electron-positron pair production (which reduces their Lorentz factor without changing their mass). Photopion production of heavy nuclei is only sizable for Lorentz factors larger than $4\times 10^{10}$, and hence is relevant only for energies larger than those considered here. 
 
 We collect all of the leading fragments heavier than H that result from the photodisintegration of a given primary element  in the mass group of that element, while the secondary protons are considered separately (the emitted neutrons will quickly decay into protons).  In this way, it is possible to introduce an effective attenuation factor for each mass group. Note that some of the leading fragments from heavy nuclei may be light, but the resulting mass distribution of the leading fragments is however generally peaked close to the mass of the primary. The total spectrum can then be obtained by adding up the contributions from the different mass groups as well as the secondary protons.
On the other hand, when computing the average logarithmic mass and its dispersion we use  the actual mass distribution of the leading fragments obtained in the simulations, since neglecting the spread in each mass group could lead to slight differences in the results. For these computations we follow \cite{hmrheavy}, using the photodisintegration cross sections from \cite{psb,salomon} and the redshift evolution of the extragalactic background light (EBL)  from \cite{in13}.

\begin{figure}[h]
\centering
\includegraphics[scale=.82,angle=0]{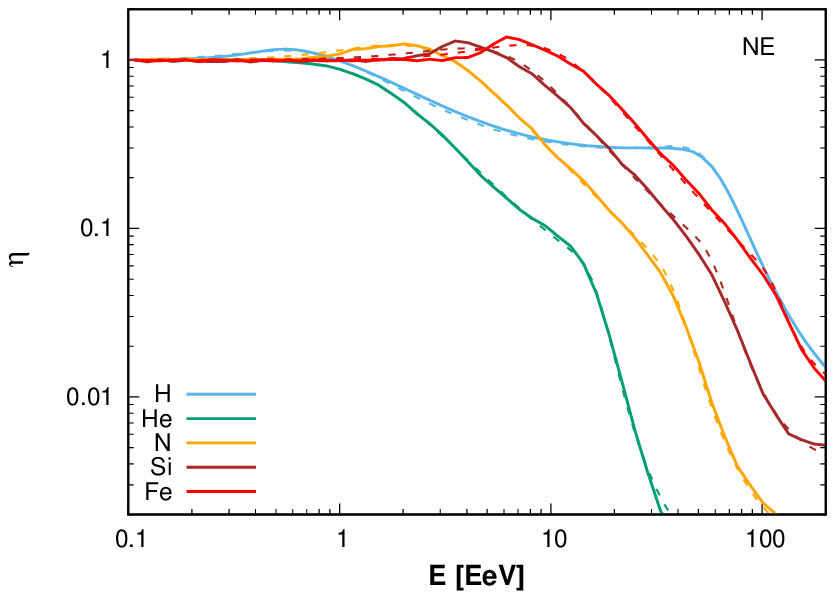}
\includegraphics[scale=.82,angle=0]{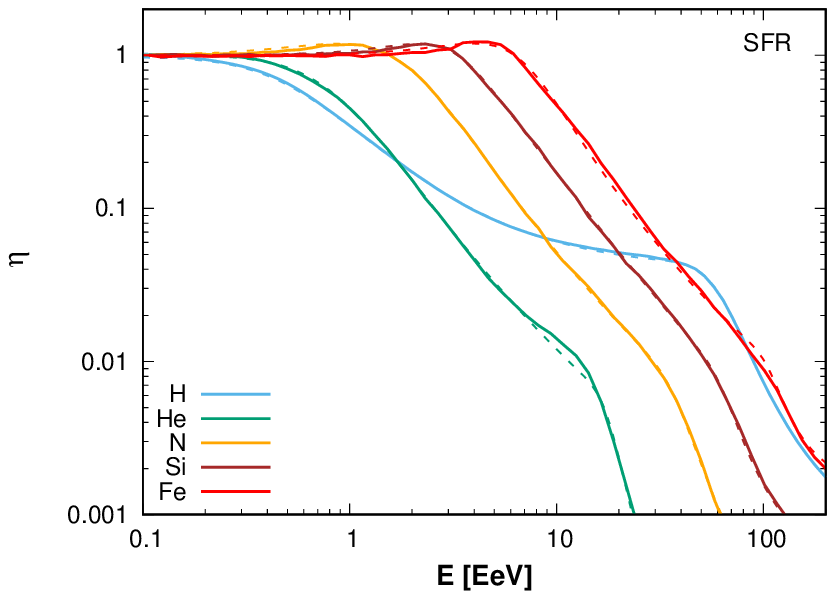}
\caption{Attenuation factor  $\eta^j(E)$ for different  primaries and for the two source evolution models. Dots are the results of the simulations and the lines are the fits obtained.}
\label{fig:etanesfr}
\end{figure}

 We show in the left panel of Figure~\ref{fig:etanesfr} the results for the case of no source evolution and in the right panel those for the SFR evolution case. The five representative mass groups are shown in both cases. The relatively larger suppression of the flux at high energies in the SFR evolution scenario is actually due to the increased luminosity of high redshift sources leading to a larger flux at low energies. Solid lines correspond to the results obtained in numerical simulations, while the dashed lines correspond to the fitted functions reported below.
 
\subsection*{Protons}

The attenuation factor for the  protons can be parametrized as
\begin{equation}
    \eta^{\rm H}(E)=\left[1/g_0(E)+ 1/g_1(E)+1/g_2(E)\right]^{-1},
\end{equation}
where the function $g_0$ accounts for the pile-up appearing at energies below the threshold of the interactions and is parametrized as
\begin{equation}
    g_0(E)\equiv (\cosh(a\, E/{\rm EeV}))^b.
\end{equation} 
The function $g_1$ accounts for the effects of the photopion production interactions while $g_2$ for those of pair production (both with the CMB). They are parametrized  in terms of the function
\begin{equation}
    F_{[A,B,C]}(E)\equiv A\exp(B\,(E/{\rm EeV})^C).
\end{equation} 

The attenuation factors for the two source evolution models considered  are then obtained from the functions 

 - no evolution (NE)
\begin{eqnarray}
g_0(E)&=& (\cosh(1.9\, E/{\rm EeV}))^{0.48}, \\
g_1(E)&=&F_{[0.0037,333,-1.03]}(E),\\
g_2(E)&=&F_{[0.24,2.2,-0.96]}(E)+F_{[0.0089,0.074,0.89]}(E).
\label{etapnoev}
\end{eqnarray}

 - star formation rate (SFR)
 \begin{eqnarray}
g_0(E)&=& 1, \\
g_1(E)&=&F_{[0.00048,515,-1.12]}(E),\\
g_2(E)&=&F_{[0.0035,5.0,-0.33]}(E)+F_{[0.001,3.2,0.021]}(E).
\label{etapsfr}
\end{eqnarray}

\subsection*{Nuclei}

The attenuation factor for the four mass groups, $j={\rm He}$, N, Si and Fe, can be parametrized with the function
\begin{equation}
    \eta^j(E)=\left[1/g^j_0(E)+ 1/g^j_1(E)+1/g^j_2(E)\right]^{-1},
    \label{eq:etaj}
\end{equation}
where now the different functions are $g^j_0(E)\equiv (\cosh(a^j\, E/{\rm EeV}))^{b^j}$ and $g^j_i(E)=F_{[A_i^j,B_i^j,C_i^j]}(E)$ for $i=1,2$. The functions $g_1^j$ account mostly for the effects of the photodisintegrations off the CMB while $g_2^j$ for those of the photodisintegrations  with the EBL, although the subdominant pair production effects are also included in them. 
The resulting coefficients of the fits are collected in Table~\ref{tab:fitnuclei}.

\begin{table}[ht!]
\centering
\caption{Coefficients of the fits to the attenuation factors for the different nuclei and for the two models of source luminosity evolution. }
\bigskip
\begin{tabular}{c c c c c c c c c c}
\hline\hline
  Evolution &  Element & $a^j$ &$b^j$ & $A^j_1$ & $B^j_1$ & $C^j_1$ & $A^j_2$& $B^j_2$ & $C^j_2$ \\
 \hline
NE & He & 0 & 1 & $8.3\times 10^{-4}$ & $2.0\times 10^{3}$&-2.1 &$7.9\times 10^{-3}$ & 6.9 & -0.43\\
 & N &1.46 & 0.36 & $1.2\times 10^{-3}$  & $6.3\times 10^{3}$ &-1.9 &$1.8\times 10^{-10}$ &24.5 &-0.062 \\
 & Si & 0.57 & 0.17 & $4.2\times 10^{-3}$  & $8.7\times 10^{4}$ &-2.4 & $9.5\times 10^{-3}$ & 13.1 & -0.45 \\
 & Fe & 0.18 &1.13 &$2.6\times 10^{-2}$ &$1.2\times 10^{11}$ & -5.2 & $1.1\times 10^{-8}$& 22.9 & -0.084\\
 \hline
SFR & He & 0 & 1 & $4.1\times 10^{-5}$ & $2.0\times 10^{3}$&-2.0 &$3.8\times 10^{-5}$ & 10 & -0.24\\
 & N & 4.5 & 0.089 & $1.2\times 10^{-4}$  & $1.4\times 10^{3}$ &-1.5 &$2.1\times 10^{-5}$ &11 &-0.21 \\
 & Si & 0.13 & 20 & $7.7\times 10^{-4}$  & $1.4\times 10^{5}$ &-2.5 & $2.6\times 10^{-17}$ & 41 & -0.047 \\
 & Fe & 0.059 & 16 & $2.9\times 10^{-3}$ &$2.7\times 10^{8}$ & -3.9 & $1.3\times 10^{-4}$& 15 & -0.27\\
\end{tabular}
\label{tab:fitnuclei}
\end{table}

\subsection*{Secondary protons}

Secondary protons get produced in significant amounts (comparable in some cases to the primary fluxes) in the energy range  between  0.1 and few EeV. Their flux depends on the source spectral index and on the cosmological source evolution considered. Their maximum energies  are actually directly related to the maximum energies of the primaries as $E_{\rm max}^{\rm sp}=E_{\rm max}^j/A\simeq E_{\rm cut}/2$. After the secondaries get produced and until they arrive to the Earth, the proton energies get degraded, mostly due to pair production and to adiabatic redshift losses. The density of secondary protons can be approximately fitted as \cite{mr19}
\begin{equation}
    \Phi^{ I}_{\rm sp}(E)\simeq \Phi^{ I}_0\sum_j f^{ I}_j \left(\frac{E}{\rm EeV}\right)^{-\gamma_{ I}}\frac{A^{2-\gamma_{ I}}g(E)}{\cosh(2E/E^{ I}_{\rm cut})},
    \label{secflux}
\end{equation}
where for  no evolution we obtain
\begin{equation}
g_{\rm NE}(E)\simeq \frac{1}{1.1 (E/{\rm EeV})^{0.75}+0.45/(E/{\rm EeV})^{1.6}},
\label{genoev}
\end{equation}
and for SFR evolution  we obtain
\begin{equation}
g_{\rm SFR}(E)\simeq \frac{1}{2.7 (E/{\rm EeV})^{1.1}+0.15/(E/{\rm EeV})^{1.4}}.
\label{gesfr}
\end{equation}

\section*{Acknowledgments}
This work was supported by CONICET (PIP 2015-0369) and ANPCyT (PICT 2016-0660). We thank the Auger Collaboration for making data available at www.auger.org.

\end{document}